\documentclass{pasa}%

\def\pasa{Publications of the Astronomical Society of Australia}

\def\Curtin{$^{1}$}
\def\CAASTRO{$^{2}$}
\def\ASU{$^{6}$}
\def\ANU{$^{7}$}

\def\CASS{$^{15}$}
\def\CfA{$^{5}$}
\def\Haystack{$^{8}$}
\def\MIT{$^{10}$}
\def\NRAO{$^{17}$}
\def\RRI{$^{9}$}
\def\SKASA{$^{3}$}
\def\Tata{$^{16}$}
\def\UMelbourne{$^{19}$}
\def\UMichigan{$^{14}$}
\def\UW{$^{11}$}
\def\UWisc{$^{13}$}
\def\Victoria{$^{12}$}
\def\SKA{$^{18}$}
\def\Rhodes{$^{4}$}

\title[High Time Resolution with the MWA]{The High Time and Frequency Resolution Capabilities of the Murchison Widefield Array}
\author[Tremblay et al.]{S.~E.~Tremblay\Curtin$^,$\CAASTRO\thanks{steven.tremblay@curtin.edu.au}, S.~M.~Ord\Curtin$^,$\CAASTRO, N.~D.~R.~Bhat\Curtin$^,$\CAASTRO, S.~J.~Tingay\Curtin$^,$\CAASTRO, B.~Crosse\Curtin, D.~Pallot\Curtin, S.~I.~Oronsaye\Curtin$^,$\CAASTRO,
G.~Bernardi\SKASA$^,$\Rhodes$^,$\CfA,
J.~D.~Bowman\ASU, 
F.~Briggs\ANU,
R.~J.~Cappallo\Haystack, 
B.~E.~Corey\Haystack, 
A.~A.~Deshpande\RRI, 
D.~Emrich\Curtin,
R.~Goeke\MIT,
L.~J.~Greenhill\CfA,
B.~J.~Hazelton\UW, 
M.~Johnston-Hollitt\Victoria,
D.~L.~Kaplan\UWisc, 
J.~C.~Kasper\UMichigan$^,$\CfA, 
E.~Kratzenberg\Haystack, 
C.~J.~Lonsdale\Haystack, 
M.~J.~Lynch\Curtin, 
S.~R.~McWhirter\Haystack,
D.~A.~Mitchell\CASS$^,$\CAASTRO, 
M.~F.~Morales\UW, 
E.~Morgan\MIT, 
D.~Oberoi\Tata, 
T.~Prabu\RRI, 
A.~E.~E.~Rogers\Haystack, 
A.~Roshi\NRAO, 
N.~Udaya~Shankar\RRI, 
K.~S.~Srivani\RRI, 
R.~Subrahmanyan\RRI$^,$\CAASTRO, 
M.~Waterson\Curtin$^,$\ANU$^,$\SKA,
R.~B.~Wayth\Curtin$^,$\CAASTRO, 
R.~L.~Webster\UMelbourne$^,$\CAASTRO, 
A.~R.~Whitney\Haystack, 
A.~Williams\Curtin \and
C.~L.~Williams\MIT
\\
$^{1}$International Centre for Radio Astronomy Research, Curtin University, Bentley, WA 6102, Australia\\
$^{2}$ARC Centre of Excellence for All-sky Astrophysics (CAASTRO)\\
$^{3}$SKA SA, 3rd Floor, The Park, Park Road, Pinelands, 7405, South Africa\\
$^{4}$Department of Physics and Electronics, Rhodes University, PO Box 94, Grahamstown, 6140, South Africa\\
$^{5}$Harvard-Smithsonian Center for Astrophysics, Cambridge, MA 02138, USA\\
$^{6}$School of Earth and Space Exploration, Arizona State University, Tempe, AZ 85287, USA\\
$^{7}$Research School of Astronomy and Astrophysics, Australian National University, Canberra, ACT 2611, Australia\\
$^{8}$MIT Haystack Observatory, Westford, MA 01886, USA\\
$^{9}$Raman Research Institute, Bangalore 560080, India\\
$^{10}$Kavli Institute for Astrophysics and Space Research, Massachusetts Institute of Technology, Cambridge, MA 02139, USA\\
$^{11}$Department of Physics, University of Washington, Seattle, WA 98195, USA\\
$^{12}$School of Chemical \& Physical Sciences, Victoria University of Wellington, PO Box 600, Wellington 6140, New Zealand\\
$^{13}$Department of Physics, University of Wisconsin--Milwaukee, Milwaukee, WI 53201, USA\\
$^{14}$Department of Atmospheric, Oceanic and Space Sciences, University of Michigan, Ann Arbor, MI 48109, USA\\
$^{15}$CSIRO Astronomy and Space Science (CASS), PO Box 76, Epping, NSW 1710, Australia\\
$^{16}$National Centre for Radio Astrophysics, Tata Institute for Fundamental Research, Pune 411007, India\\
$^{17}$National Radio Astronomy Observatory, Charlottesville and Greenbank, USA\\
$^{18}$SKA Organization, Jodrell Bank Observatory, Lower Withington, Macclesfield, SK11 9DL, United Kingdom\\
$^{19}$School of Physics, The University of Melbourne, Parkville, VIC 3010, Australia\\
}

\jid{PASA}
\doi{10.1017/pas.\the\year.xxx}
\jyear{\the\year}

\usepackage{comment} 
\usepackage{caption}
\usepackage{subcaption}
\usepackage{amssymb}
\usepackage{upgreek}
\usepackage{epstopdf}
\usepackage{color}

\newcommand{\ga}{$\gtrsim$}
\newcommand{\la}{$\lesssim$}
\newcommand{\dmu}{pc cm$^{-3}$}

\newcommand{\aj}{AJ}			
\newcommand{\araa}{ARA\&A}		
\newcommand{\apj}{ApJ}			
\newcommand{\apjl}{ApJLett}		
\newcommand{\aap}{A\&A}			
\newcommand{\mnras}{MNRAS}		
\newcommand{\nat}{Nature}		

\begin{document}%
\begin{abstract}
The science cases for incorporating high time resolution capabilities into modern radio telescopes are as numerous as they are compelling. Science targets range from exotic sources such as pulsars, to our Sun, to  recently detected possible extragalactic bursts of radio emission, the so-called fast radio bursts (FRBs). Originally conceived purely as an imaging telescope, the initial design of the Murchison Widefield Array (MWA) did not include the ability to access high time and frequency resolution voltage data. However, the flexibility of the MWA's software correlator allowed an off-the-shelf solution for adding this capability. This paper describes the system that records the 100 $ \upmu$s and 10 kHz resolution voltage data from the MWA. Example science applications, where this capability is critical, are presented, as well as  accompanying commissioning results from this mode to demonstrate verification.
\end{abstract}
\maketitle%

\section{INTRODUCTION}
\label{sec:intro}


The canonical data path through a radio interferometer includes cross-correlation, since this provides a sampling of the spatial coherence of the sky brightness distribution on scales commensurate with that of the interferometer baseline distribution. These {\em visibilities} are required in order to reconstruct the image plane from the observed data (Thompson, Moran \& Swenson 2007\nocite{2007isra.book.....T}). Since this process alone increases the data volume by a factor of $(N-1)/2$, where $N$ is the number of antennas, time averaging is invariably included into this operation as a way of increasing the signal to noise of the visibility set. This integration can continue as long as the visibilities remain coherent, and significantly reduces the data volume (typically 4-6 orders of magnitude).  Obviously, phenomena on timescales shorter than the correlation integration time (typically on the order of one to a few seconds) will be smeared or even reduced below detection through this averaging. While sources varying on these timescales (e.g. pulsars, fast transients) have typically been observed by single-dish radio telescopes (e.g. Parkes, Green Bank Telescope, Arecibo) the increase in resolution available with present-day interferometers argues strongly for their use to observe these exotic astronomical objects. Looking ahead toward the future of radio astronomy instrumentation, which is focused on even larger, more elaborate instruments such as the constituents of the Square Kilometre Array (SKA, Dewdney et al. 2013\nocite{2013SKA...Base}) where interferometric sensitivity will surpass that of single-dishes, implementing high time resolution observations in the SKA Precursor instruments seems an obvious step.

The Murchison Widefield Array (MWA; Tingay et al. 2013a\nocite{2013PASA...30....7T}; Lonsdale et al. 2009\nocite{2009IEEEP..97.1497L}) is a new low frequency (80-300 MHz) radio interferometer located, roughly 600 km north of Perth, at the CSIRO Murchison Radio-astronomy Observatory (MRO) in Western Australia which has recently entered regular operations as the low frequency precursor to the SKA. Its four key science themes are: 1) statistical detection of the Epoch of Reionisation; 2) Galactic and extragalactic astrophysical processes; 3) Solar, heliospheric and ionospheric imaging and analysis; and 4) time domain astrophysics (Bowman et al. 2013\nocite{2013PASA...30...31B}).  The standard MWA signal path typically generates visibility cubes from the correlator at a cadence of 500~ms from 24 $\times$ 1.28 MHz wide bands with a frequency resolution of 40 kHz. Details of the correlator implementation are described at length in Ord et al. (2014\nocite{Ord...submitted}). While there exists some flexibility in the integration time and spectral averaging within the correlator, the data rate must be limited by what can be sustainably transferred into the archive.

This paper describes a new system, which enables recording of the entire channelised voltage input to the correlator, preserving the full instrumental resolution in both the time and frequency domains. This system is called the Voltage Capture System (VCS). We also present example science use cases, for which the VCS capability is critical, together with the results of commissioning observations that demonstrate performance in these science areas. The VCS mode was offered for the first time as part of the 2015-A MWA observing semester (January - June, 2015) and five proposals that focused on this mode were received, ranging from solar science to FRB and pulsar studies, to instrument verification, confirming the demand for such a system at the MWA.

\section{VOLTAGE CAPTURE SYSTEM}
\subsection{Data Capture}
The standard MWA signal path, as described in detail in Tingay et al. \shortcite{2013PASA...30....7T}, Prabu et al. (2014\nocite{Prabu...submitted}) and Ord et al. (2014\nocite{Ord...submitted}), can be simplified to the following synopsis. Dual-polarisation dipole antennas arranged as 4\,x\,4 grids (tiles) are analogue beamformed, with the radio frequency signals sent to receiver boxes in the field. Each of the 256 signals (128 tiles x 2 polarisations across 16 receivers) is subsequently amplified, digitised and processed through a coarse polyphase filter bank (PFB) in the receiver enclosures. The outputs from each receiver are then sent to one of four fine PFBs (1/4 of the tiles to each). The 32 fine PFB outputs are then converted from RocketIO (a Xilinx serial protocol) to Transmission Control Protocol (TCP) by 16 media converter servers (CISCO UCS C240) using Peripheral Component Interconnect Express mounted cards produced by Engineering Design Team Incorporated (EDT Cards). These 32 signals are then reorganised into coarse channel groups and sent to the 24 servers hosting the Graphics Processing Unit (GPU) cards that perform the cross-multiply step of correlation. After correlation, the output is transferred offsite and archived.

\begin{figure*}
\begin{center}
\includegraphics[width=32pc, height=20pc]{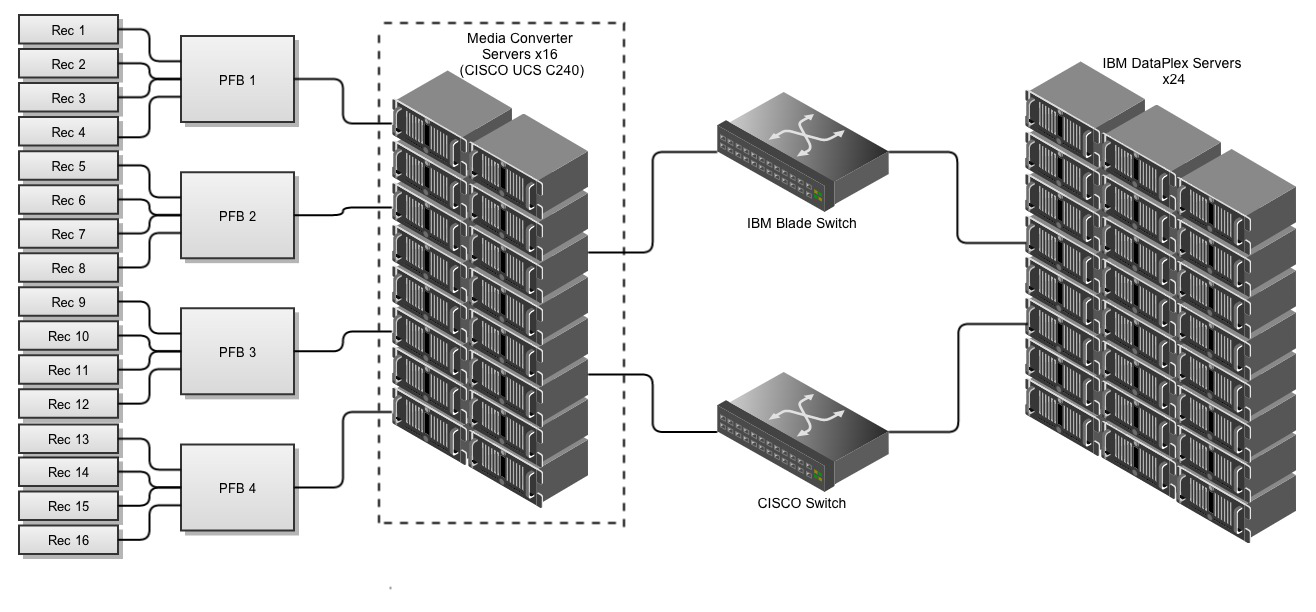}
\caption{Simplified illustration of the MWA digital signal path. For the VCS mode, the baseband data between the fine channel polyphase filter bank (PFB) and the X-engine of the correlator are written to local RAIDs on the 16 media converter servers (highlighted with a dashed box). This gives us 100 $ \upmu$s resolution and frequency channels 10 kHz wide.}
 \label{fig:digital_path}
\end{center}
\end{figure*}

The initial digital system design did not include any access to the voltage data along this path. We decided to implement voltage capture on the media conversion servers since they otherwise have a relatively small workload and had not been purchased or even specified at the time of this subsystem's design. The media converter servers were subsequently specified to include 128 GB of RAM each, for potential buffering purposes, 24 $\times$ 2.5" HDD slots, and a dual-channel drive controller. The digital signal path, receivers through GPU servers, are schematically represented in Figure \ref{fig:digital_path} to highlight where the voltages are recorded. 

Each server is equipped with two redundant arrays of independent disks (RAIDs) for recording voltage data. These are each comprised of six 2.5" 300 GB (279 GB actual) 10K SAS drives and are combined as a hardware controlled RAID 0 (block level striping but without parity or mirroring), giving each RAID 1.44 TB of usable storage with no redundancy. The RAIDs are on independent 6 Gb/s channels in the controller, to maximise data throughput.

The 4-bit + 4-bit complex voltages from the fine PFB stream through the media conversion servers at a rate of $\sim$ 4.2 Gbps. After being converted by the EDT cards, these data can be streamed directly to the RAIDs. Each server has two PFB lanes passing through it, each of which is directed to a separate RAID in the unaltered post-PFB format as files on 1 second boundaries. Therefore, every second of VCS observation generates 32 $\times$ 242 MB files across 16 machines (where each file contains 1/8 of the fine channels for 1/4 of the tiles) for a combined rate of 7.744 GB/s.

It is important to highlight that the `original' digitised tile voltages are not recorded, instead it is the channelized output of the four PFBs (i.e. the data have been aggregated, filtered, amplified and processed through two PFBs before being recorded). 

\subsection{Capabilities}
\label{sec:capabilities}
VCS Observations consist of 30.72 MHz of bandwidth, in 1.28 MHz wide subbands, spread as desired between 80 and 300 MHz. These data have 10 kHz channels and 100 $ \upmu$s time resolution (corresponding to critically sampled 10-kHz channels at the Nyquist-Shannon rate). These data are the same as the standard correlator input. At most, the system can record a single uninterrupted time span of $\sim$ 100 minutes. Dedicated hardware, in the form of a server with 16 $\times$ 3.5" SAS-drive bays, is installed in Perth where the recorded data can be transferred onto a software RAID utilising a 10 Gb/s link to the MRO. Significant time, however,  is required between subsequent VCS observations to transport the data from the site since archiving correlated observations takes precedence.

Recording the data in this mode gives the user maximum flexibility in processing the data, in particular how the 256 data streams (polarisations $\times$ tiles) are combined. The simplest merger of these data is the incoherent sum, where the voltages from each tile are multiplied by their complex conjugate, to form the power,  and subsequently summed. This preserves the field of view (FoV) while increasing the sensitivity over a single tile by a factor of $\sqrt{N}$, where $N$ is the number of tiles summed. All the data presented in this paper were combined in this fashion unless noted otherwise.

Alternately, more than an order of magnitude increase in sensitivity can be gained if a phase rotation is applied to each voltage stream before summing the voltages to form a coherent beam at the cost of a drastic reduction in the field of view for an individual phase centred beam. To attain the same sky coverage as the incoherent sum, thousands of coherent beams need to be processed. Since these data are recorded from the `standard' data path, it is also possible to cross-correlate these data offline in a manner similar to normal but with control over the temporal and spectral integration (within the constraints of the raw data and the available compute resources).

\section{VCS COMMISSIONING RESULTS}

As the VCS mode was added to the MWA, a wide range of engineering tests and on-sky commissioning data were taken to verify the elements as they were added as well as for data pipeline development. These data, taken with a variety of bandwidths and number of tiles, are presented here to demonstrate the capabilities of the instrument. VCS observations in this paper are labeled \textit{YYYYMMDDX} where \textit{YYYY} is the year, \textit{MM} is the month, \textit{DD} is the day of each observation and \textit{X} is an incremental letter denoting the order of observations for the day (i.e. A, B, C, ...).

\subsection{Pulsar Science}
\label{sec:pulsar}

Arguably the primary science application of the voltage capture mode is for pulsar observations. Following their serendipitous discovery at 81.5 MHz by Hewish et al. \shortcite{1968Natur.217..709H},  much of the early research on pulsars was at low frequencies (Taylor \& Manchester 1977\nocite{1977ARA&A..15...19T}), however the eventual quest to find more pulsars in the Galactic plane, which is highly sky-noise dominated at low frequencies, and also to achieve higher precision in timing pushed the observations to higher frequencies (\ga 1 GHz). With the advent of multiple new low-frequency arrays including MWA, the Low Frequency Array (LOFAR; van Haarlem et al. 2013\nocite{2013A&A...556A...2V}), and the Long Wavelength Array (LWA; Taylor et al. 2012\nocite{2012JAI.....150004T}), a renaissance in low-frequency pulsar astrophysics is on the horizon. In fact, a number of low-frequency pulsar results have already started being published from these and other instruments (e.g. Bhat et al. 2014\nocite{2014ApJ...791L..32B}, Archibald et al. 2014\nocite{2014ApJ...790L..22A}, Dowell et al. 2013\nocite{2013ApJ...775L..28D}, Stovall et al. 2014\nocite{2014ApJ...791...67S}).

\begin{figure}
\includegraphics[width=20pc]{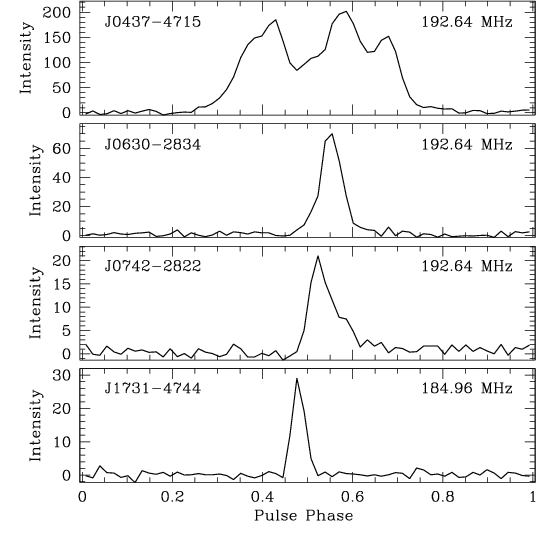}
\caption{Pulse profiles of four different pulsars observed with the VCS mode covering a range of DMs and periods (see Table \ref{tab:pulsars} for values) to demonstrate the flexibility of the instrument. There are 64 bins, spanning the pulse period for each pulsar.}
\label{fig:profiles}
\end{figure}

While state-of-the-art pulsar backends are capable of providing far superior time resolution via phase-coherent de-dispersion over large bandwidths (e.g. van Straten \& Bailes 2011\nocite{2011PASA...28....1V}), the VCS functionality of the MWA is well matched to a wide range of pulsar science goals at low frequencies, particularly for long-period pulsars (spin period, $P$ \ga 100 ms) with dispersion measures (DMs) \la 200 \dmu, where scattering is not large enough to significantly smear the emission across the pulse period. This parameter space samples the vast majority of the local (\la 1 kpc) pulsar population. Our 10 kHz channelization means the effective achievable time resolution will largely be limited by the dispersive smearing within the channel, which is $\sim$ 1 ms at DM = 100 \dmu at a frequency of 200 MHz. In targeted pulsar observations, this limitation is greatly reduced by coherent dedispersion. The 100 $ \upmu$s resolution presents an obvious challenge in observing millisecond pulsars (MSPs).  It is, however, enough resolution to construct high-quality pulse profiles of pulsars with periods $\ge$ that of PSRJ0437-4715 (see  Bhat et al. 2014 and Fig. \ref{fig:profiles}). It is, in principle, possible to reconstruct the time resolution of the full bandwidth ($\sim$32.6 ns)by inverting the second PFB stage, although this is yet to be attempted.

Our successful detection of PSR J0437$-$4715, a binary MSP with P=5.75 ms and DM=2.65 \dmu, also provides an excellent demonstration of the timing stability of our recording system. Multiple observations of this pulsar and other objects such as the Crab pulsar have been made over time durations of 1 hr without encountering any recording glitches. A close inspection of commissioning data (Table \ref{tab:pulsars}) confirms that, barring some issues such as power level modulations caused by the Orbcomm satellites within the FoV (see Section \ref{sec:rfi}) and abrupt changes in power levels that result from changes in the beam-former settings, the data are in general highly stable. While we are unable to accurately determine the exact instrumental offset in our timing owing to our limited data, we have verified our time stamping accuracy, to first order, by successfully combining multiple observations of PSRs J0630$-$2834 and J0534+2200 that span time intervals in the range of $\sim$30 minutes to 20 days (Table \ref{tab:pulsars}). The pulse profiles from different observations align at a level suggesting that the offset is limited to \la0.5 ms. We aim to further characterise this more accurately as more observations accrue over the course of time.

Pulse broadening resulting from multipath scattering in the interstellar medium (ISM) is an important consideration at the MWA's frequencies. Based on the Galactic electron density models (NE2001; Cordes \& Lazio 2002\nocite{2002astro.ph..7156C}) and an observationally-established scaling relation between scattering, DM, and the observing frequency (Bhat et al. 2004\nocite{2004ApJ...605..759B}), a scatter broadening $\tau_{scatt} $ $\sim$ 1 ms is expected  near the low end of the MWA band ($\sim$ 100 MHz) toward pulsars at distances $\sim$1 kpc located within the Galactic plane ($b=0^{\circ}$). At larger distances within the plane, scattering can be substantial, owing to the non-linear scaling of scattering with DM, with $\tau_{scatt}$ $\sim$100 ms to be expected toward the Galactic centre at a distance of $\sim$3 kpc at $\sim$100 MHz. Off the Galactic plane, scattering will likely be far less severe, and is expected to be \la 0.3 ms at frequencies \ga 200 MHz, thereby retaining sensitivity to the detection of most nearby pulsars, including MSPs.  

As seen from Table \ref{tab:pulsars} and Fig. \ref{fig:profiles}, our commissioning analysis so far investigated DMs up to 123 \dmu, and little scattering is evident at DMs \la 50 \dmu at $\sim$200 MHz. Besides the Crab, known for its atypical scattering (Ellingson et al. 2013\nocite{2013ApJ...768..136E}; Bhat et al. 2007\nocite{2007ApJ...665..618B} ), PSR J0742-2834 is the only pulsar that shows scattering in our data (Fig. \ref{fig:profiles}). While generally considered to be a hindrance for most pulsar studies including searches, scattering measurements provide useful means of characterising the ISM turbulence, for which the MWA frequency band is well suited.  

Even with the 100-$ \upmu$s, 10-kHz limitations of the VCS, useful scintillation and profile studies are possible even for MSPs, as vividly demonstrated in Bhat et al. \shortcite{2014ApJ...791L..32B}. The 100-$ \upmu$s resolution, while not a limitation for scintillation studies where the primary goal is to investigate the
time and frequency modulations of the integrated pulse emission, will likely pose a limitation in the profile studies of MSPs with $P$ $\sim$ a few ms. Similarly, the 10-kHz resolution may limit the scope of scintillation studies to DMs \la 40 \dmu\ in the MWA band. However, cyclic spectroscopy could be implemented to characterise scattering at higher DMs (Demorest 2011\nocite{2011MNRAS.416.2821D}). Low-frequency observations can also potentially yield accurate DM measurements, provided the frequency-dependent evolution of the pulse profile is modelled in the analysis; this is important for timing-array applications such as the search for gravitational waves (e.g. Manchester et al. 2013\nocite{2013PASA...30...17M}). 

\begin{table*}
\caption{Pulsars detected after processing MWA voltage data. The voltage streams were combined into incoherent beams and then processed using a PRESTO pipeline. These observations were performed at various times within the commissioning period of the VCS, hence the variety of bandwidths. PSRs J0534+2200 (the Crab) and J0528+2200 were detected within the same beam during an observation, similarly PSRs J0630-2834 and J0742-2822 within a single 31 minute observation.}
\begin{center}
\begin{tabular}{@{}cccccccc@{}}
\hline\hline
Pulsar & Period$^1$  & DM $^1$ & MJD & Centre Frequency & Bandwidth & Dwell & S/N \\
&  (ms) & (\dmu)& &(MHz) & (MHz) & (min) &   \\
\hline%
J0437-4715 & 5.75 & 2.64& 56559 & 192.64 & 15.36 & 60 & 202 \\
 &  & & 56639 & 192.64 & 15.36 & 60 & 85 \\
J0528+2200 & 3745  & 50.94& 56559 & 192.64 & 15.36 & 35& 9 \\
J0534+2200 & 33 & 56.76& 56558 & 192.64 & 15.36 & 44 & 108 \\
&  & & 56559 & 192.64 & 15.36 & 35 & 133\\
J0630-2834 & 1244 & 34.50& 56596 & 192.64 & 15.36 & 23  & 54 \\
 &  & & 56615 & 192.64 & 15.36 & 10 & 70 \\
&  &  & 56615 & 192.64 & 15.36 &  13 & 36 \\
J0742-2822 & 166.8 & 73.95& 56615 & 192.64 & 15.36  & 20 & 24 \\
J0953+0755 & 253 & 2.96 & 56192 & 155.52 & 1.28 & 10 & 32$^\dagger$ \\
 & & & 56538 & 147.2 & 14.08$^\ddagger$ & 13 & 14 \\
J1136+1511 & 1188 & 4.85& 56540 &147.2 & 15.36 & 28 & 18 \\
J1731-4744 & 830 & 123& 56880 &184.96 & 30.72 & 64 & 29 \\
J1752-2806 & 562 & 50.37& 56192 &151.8 & 1.28 & 15 & 6$^\dagger$ \\
J1921+2153 & 1337 & 12.44 & 56923 & 184.96 & 30.72 & 65 & 136 \\
\hline\hline
\end{tabular}
\end{center}
\label{tab:pulsars}
\tabnote{$^\dagger$ These observations were done with only 32 tiles.}
\tabnote{$^\ddagger$ The bandwidth spanned 15.36 MHz, but one of the central course channels was not recorded so only 14.08 MHz of data were summed in the dedispersed time series.}
\tabnote{$^1$ All values are taken from the ATNF Pulsar catalogue http://www.atnf.csiro.au/people/pulsar/psrcat/ (Manchester et al. 2005\nocite{2005AJ....129.1993M}) }
\end{table*}

\subsection{Single Pulse Astrophysics}
\label{sec:single_pulse}
The last ten years has seen renewed interest in high time resolution single pulse detections, with particular enthusiasm for both rotating radio transients (RRATs; McLaughlin et al. 2006\nocite{2006Natur.439..817M}, Burke-Spolaor et al. 2011\nocite{2011MNRAS.416.2465B}, Keane et al. 2011\nocite{2011MNRAS.415.3065K}) and fast radio bursts (FRBs; Lorimer et al. 2007\nocite{2007Sci...318..777L}, Thornton et al. 2013\nocite{2013Sci...341...53T}, Spitler et al. 2014\nocite{2014arXiv1404.2934S}). Both the large field of view (hundreds or even thousands of square degrees) as well as the flexibility in observing frequency make low frequency arrays, such as the MWA, powerful tools in the search for single pulse emission. As mentioned in section \ref{sec:capabilities}, an incoherent sum of the elements preserves this large field of view and generates a data set that is practical to search through using modest compute resources. Once detected, the power of the interferometric array can be used to localise and study these pulses as long as the voltages have been preserved.  

\begin{figure}
\begin{center}
\includegraphics[width=22pc, height=12pc]{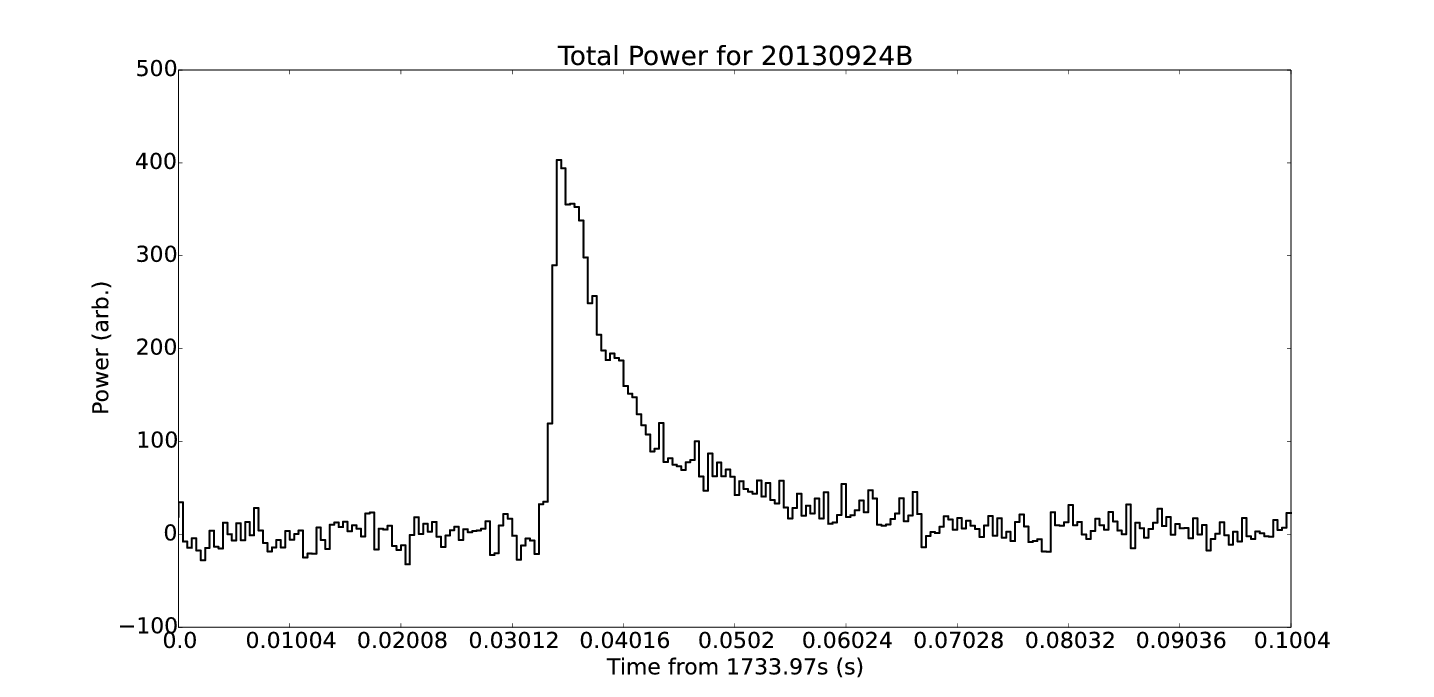}
\caption{Dedispersed (56.76 pc cm$^{-3}$) total power from one of the detected Crab giant pulses observed with the VCS at 192.64 MHz. The long scattering tail ($\sim$40 ms) extends further than a single pulse period (33 ms). The median power from each fine channel was removed before dedispersion and time steps are averaged to 400 $ \upmu$s. Reference times on the abscissa denote seconds from the beginning of the observation.}
\label{fig:cgp}
\end{center}
\end{figure}

We have verified the ability of the MWA to observe short-duration single pulse emission using two separate observations of the Crab pulsar (PSR\,J0534+2200, Table \ref{tab:pulsars}) to observe giant pulses (Hankins et al. 2003\nocite{2003Natur.422..141H}) which, based on higher frequency observations ($\sim$5 GHz), are thought to be intrinsically $\sim$nanosecond duration  bright pulses. In the MWA band, these intrinsically short pulses are scattered to widths of tens of milliseconds ( see Fig. \ref{fig:cgp}). We detected 51 and 47 pulses above 6\,$\sigma$\footnote{These signal to noise measurements are for detection with a 12-ms boxcar starting from the rising edge of the pulse in an attempt to recover signal from the scattering tail. Matched filter detections of these pulses would undoubtedly recover more power and yield higher signal to noise detections.\label{fn:sigma}} from the observations. The brightest pulse we detected was 39\,$\sigma$\footnotemark[1] (Fig. \ref{fig:cgp}), and shows a well defined sharp peak followed by an exponential scattering tail. 

For further information on FRBs see Trott, Tingay \& Wayth \shortcite{2013ApJ...776L..16T} for the prospects of detecting FRBs with the MWA over a variety of potential spectral indices and Tremblay et al. \shortcite{Tremblay...inprep} which describes the FRB pipeline being used by the MWA.

\subsection{Solar Science}
\begin{figure}
\begin{center}
\includegraphics[width=24pc, height=14pc]{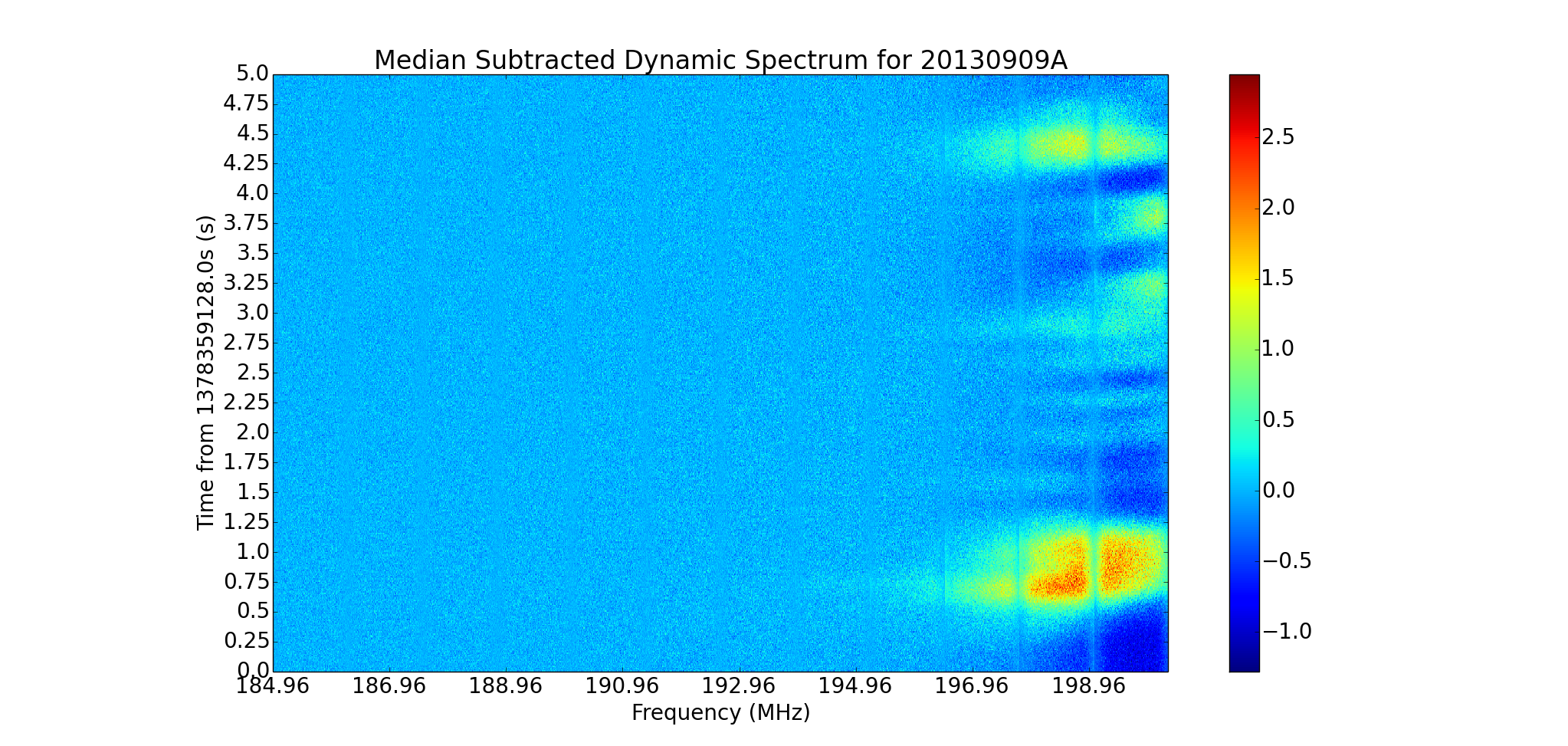}
\caption{Dynamic spectrum spanning 5 seconds where the MWA's 128 tiles have been incoherently summed from observations on MJD 56544. Here, 12 coarse channels (half of the MWA's typical bandwidth) were averaged to 10 ms increments and the median value of each 10-kHz fine channel was subsequently subtracted to highlight the variable emission. The vertical bands are the result of coarse channel edges where sensitivity is reduced. The narrowband, sub-second solar features highlighted here would be smoothed out with the typical integration times used in standard imaging modes. Times on the ordinate reference Unix Time.}
\label{fig:solar}
\end{center}
\end{figure}

The standard imaging modes of the MWA are being used to study the Sun (Oberoi et al. 2014\nocite{2014arXiv1403.6250O}) with a variety of aims, including the investigation of the quiet Sun, the detailed study of Type II and III bursts, and the characterisation of Coronal Mass Ejections (described in Bowman et al. 2013\nocite{2013PASA...30...31B}, and references therein).  However, the temporal resolution of the standard MWA imaging modes can significantly under-sample solar radio emission variability.  Intense solar bursts evolve rapidly in time and frequency, but previous MWA observations show that even in a quiet state, low level radio emission from the Sun evolves strongly in both time and frequency (Oberoi et al. 2011\nocite{2011ApJ...728L..27O}).  As such, there is a need for significantly higher time resolution observation modes with the MWA. Observations of the Sun for which voltages are captured can
be used for high time resolution beamforming, either incoherent or coherent as mentioned in section \ref{sec:capabilities}, to form high time resolution dynamic spectra (for example, as seen in Figure \ref{fig:solar}).  Furthermore, the voltages can be correlated at high time resolution
post-observation, in order to undertake high time resolution imaging of the Sun.  Both high cadence beamforming and imaging modes are identical to those to be used for pulsars and searches for fast radio bursts (sections \ref{sec:pulsar} and \ref{sec:single_pulse}).

In addition to Solar studies, the VCS is anticipated to conduct observations of interplanetary and ionospheric scintillations. High time resolution observations of satellite beacons will also be used to study the variability of ionospheric Faraday rotation.

\section{INTERFERENCE ENVIRONMENT ON SUB-SECOND TIME SCALES }
\label{sec:rfi}
Using the VCS to perform a comprehensive survey of the radio frequency interference (RFI) environment at the MRO in the MWA band is anticipated, but was outside the scope of commissioning this new mode. We have, however, inevitably detected a variety of forms of interference (satellite, airborne and ground based) throughout the commissioning process.

The  most common strong source of interference we have encountered to date is generated by the Orbcomm satellite constellation, a global communications network operating in multiple bands around 137 MHz. While these transmissions to Earth are restricted to narrow bands which are, in principle, avoidable by wise coarse channel selection, the transmitted power is so much greater than astronomical sources that this additional power saturates a portion of the signal chain, with gain values typically used for observations, and power spreads out across the wider band. The MWA digitiser is presented with 80-300 MHz and always produces a complete set of 256 coarse channels, from which 24 are selected and sent to the fine PFB (see Prabu et al. 2014\nocite{Prabu...submitted} for further details). Therefore, simply avoiding Orbcomm's broadcast frequency is not sufficient to mitigate this effect. It is worth noting that incoherently summing the tiles, as we have done here, has maximum sensitivity to this (and other) interference. Rotating the phases independently to form coherent beams, and cross-correlation, both decorrelate this emission in directions other than the actual line of sight raising the overall system temperature but reducing the coherence of the interfering emission.

\begin{figure}
\begin{center}
\includegraphics[width=22pc, height=13pc]{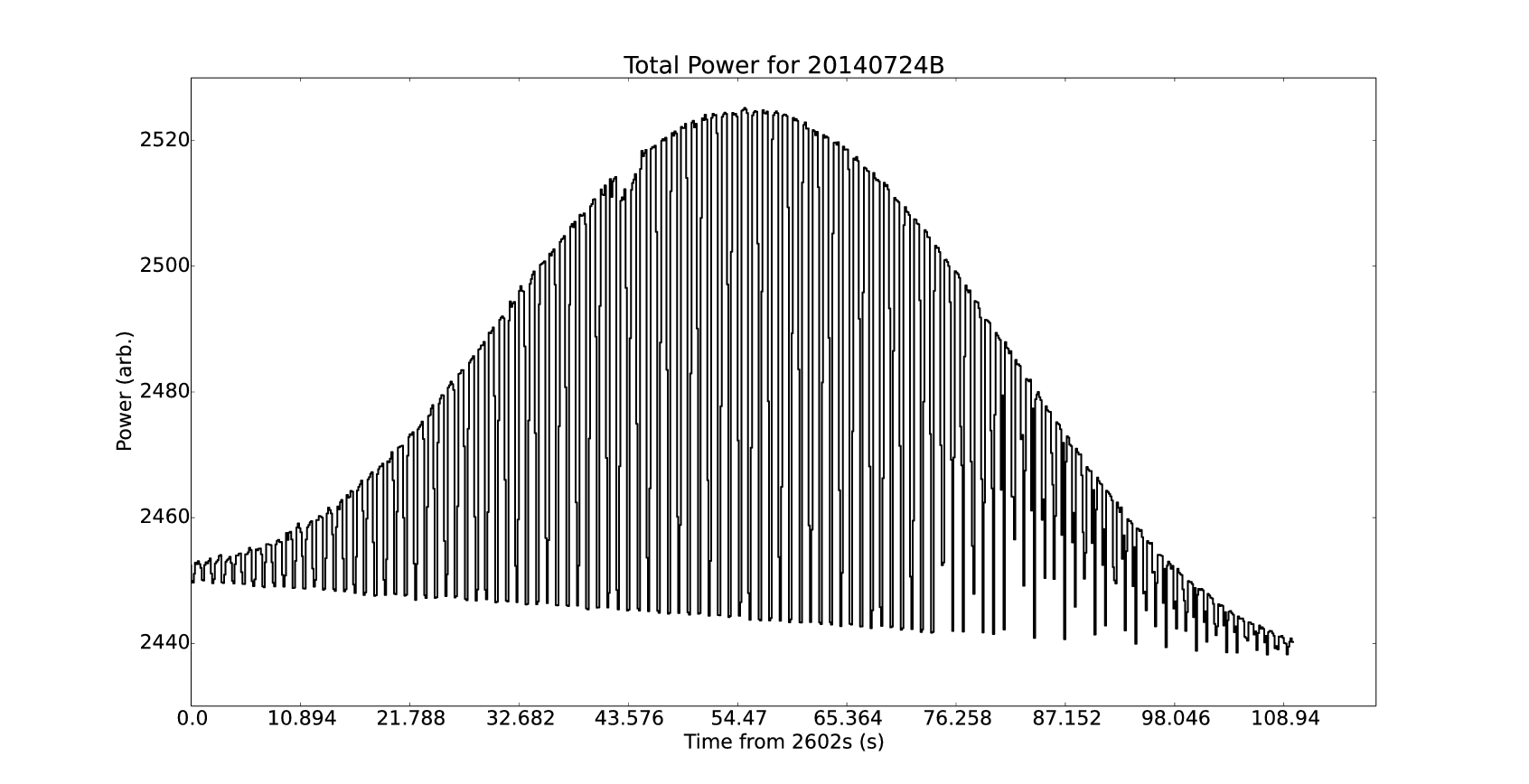}
\caption{Total power (summed over tiles and frequencies) plot showing the effect an Orbcomm satellite has on a 184.96 MHz observation when transmitting within our beam, the envelope of the signal amplitude tracing out the tile gain pattern. These data have been averaged to 130 ms. Note the 1 Hz power variations, which would cause problems for blind periodic pulsar searches. Reference time on axis is seconds from the beginning of the observation.}
 \label{fig:orbcomm_pow}
\end{center}
\end{figure}

Orbcomm transmissions are time-domain multiplexed at 1 Hz (Fig. \ref{fig:orbcomm_pow}), making this interference a complication for pulsar searching.

\begin{figure}
\begin{center}
\includegraphics[width=23.5pc, height=14pc]{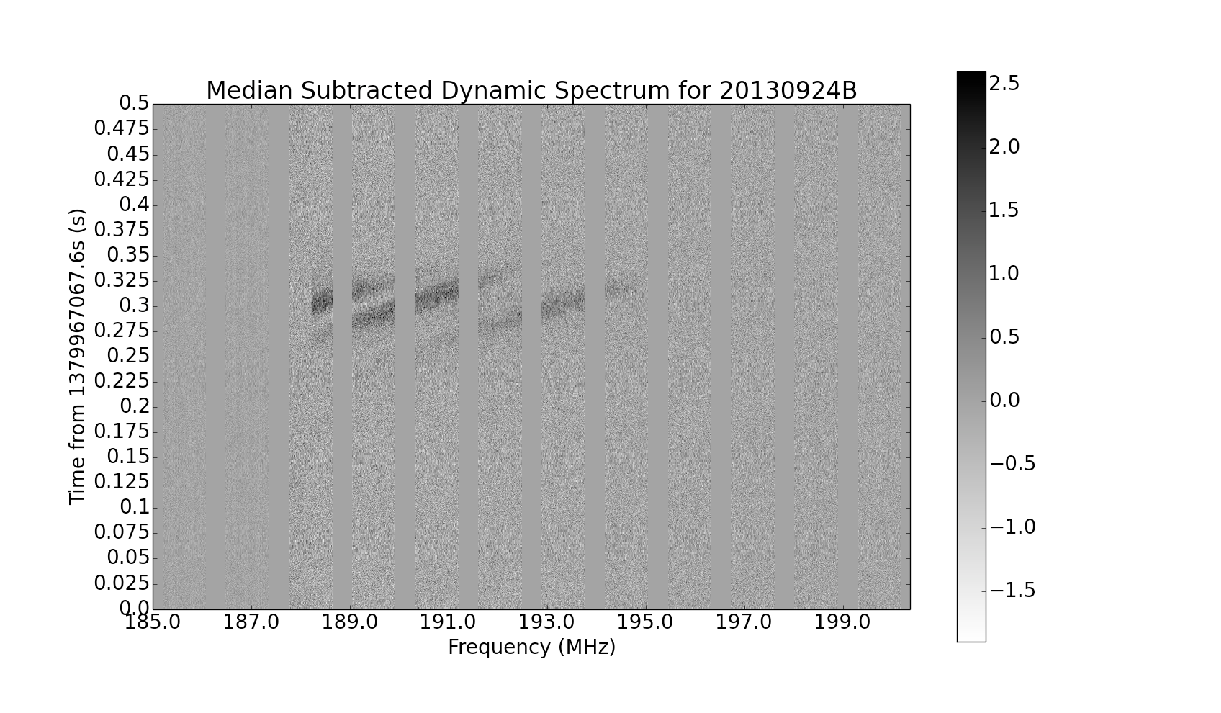}
\caption{A dynamic spectrum from MJD 56559 spanning 0.5 seconds and showing short lived multi-path interference fringes from TV Channel 8 (8 MHz centred approximately around 191.5 MHz). The median from each frequency bin was subtracted before plotting. Time is averaged to 1 ms (the plot spans 0.5 seconds) and frequency is kept at 10 kHz. Once again, the vertical lines between coarse channels are visible through the variable emission.}
 \label{fig:rfi}
\end{center}
\end{figure}

The very nature of our high time resolution observations means that we are also able to detect short duration terrestrial RFI that normally goes unnoticed due to averaging, even in an RFI environment as quiet as the MRO. These detections often come about due to short-lived multi-path propagation involving the atmosphere and/or ionosphere causing temporary enhancements in communications signals (TV, FM, inter aircraft, etc.; Fig \ref{fig:rfi}).

The redirected RFI seen by the MWA has been shown to be a useful tool for detection of near Earth objects such as the Moon (McKinley et al. 2013\nocite{2013AJ....145...23M}) and low-orbiting space debris (Tingay et al. 2013b\nocite{2013AJ....146..103T}).

While not a broadcast form of interference, the other event which has affected our data are the step transitions generated by the MWA beamformers. Each MWA tile has an analogue beamformer, with a discreet set of `best value' pointings. Instead of continuously repointing, the MWA evaluates whether or not to switch between these delay settings at fixed time intervals (typically 296 s). The combination of these quantised repointings and the large field of view of the incoherent sum leads to abrupt changes in the total power. This expresses itself as a step function in a time series in incoherently summed observations. Once again, this would not affect coherent post-processing as strongly.

\section{CONCLUSIONS}

In this paper, we have introduced the capability of the MWA to record its high-time (100 $ \upmu$s) and high frequency (10 kHz) resolution post-PFB voltage stream, the Voltage Capture System. We have demonstrated the abilities of this mode by observing ten separate pulsars throughout different phases of commissioning. We have also exercised this mode on dispersed, single-pulse signals by observing the Crab pulsar and detecting roughly one giant pulse a minute with only half of the typical MWA bandwidth and only combining the tiles in an incoherent fashion. These detections, and the stability of the observations we have performed, highlight the MWA's ability to become a workhorse in these, and other, areas of radio astronomy.
 
\begin{acknowledgements}
This scientific work makes use of the Murchison Radio-astronomy Observatory, operated by CSIRO. We acknowledge the Wajarri Yamatji people as the traditional owners of the Observatory site. Support for the MWA comes from the U.S. National Science Foundation (grants AST-0457585, PHY-0835713, CAREER-0847753, and AST-0908884), the Australian Research Council (LIEF grants LE0775621 and LE0882938), the U.S. Air Force Office of Scientific Research (grant FA9550-0510247), and the Centre for All-sky Astrophysics (an Australian Research Council Centre of Excellence funded by grant CE110001020). Support is also provided by the Smithsonian Astrophysical Observatory, the MIT School of Science, the Raman Research Institute, the Australian National University, and the Victoria University of Wellington (via grant MED-E1799 from the New Zealand Ministry of Economic Development and an IBM Shared University Research Grant). The Australian Federal government provides additional support via the Commonwealth Scientific and Industrial Research Organisation (CSIRO), National Collaborative Research Infrastructure Strategy, Education Investment Fund, and the Australia India Strategic Research Fund, and Astronomy Australia Limited, under contract to Curtin University. We acknowledge the iVEC Petabyte Data Store, the Initiative in Innovative Computing and the CUDA Center for Excellence sponsored by NVIDIA at Harvard University, and the International Centre for Radio Astronomy Research (ICRAR), a Joint Venture of Curtin University and The University of Western Australia, funded by the Western Australian State government. This research was conducted by the Australian Research Council Centre of Excellence for All-sky Astrophysics (CAASTRO), through project number CE110001020.The National Radio Astronomy Observatory is a facility of the National Science Foundation operated under cooperative agreement by Associated Universities, Inc. DLK was partially funded by NSF grant AST-1412421.

\end{acknowledgements}

\end{document}